\documentclass[preprint,amsmath,amssymb,aps,showkeys,pre]{revtex4-2}
\usepackage[english]{babel}
\usepackage{color}
\usepackage{amscd}
\usepackage{epsfig}
\usepackage{tabularx}
\usepackage{graphicx}
\usepackage{latexsym}
\usepackage{amsmath}
\usepackage{amsfonts}
\usepackage{amssymb}
\usepackage[latin1]{inputenc}
\usepackage{times}
\usepackage[T1]{fontenc}
\usepackage{latexsym}
\usepackage{graphics}
\usepackage{verbatim}
\usepackage[absolute]{textpos}
\usepackage{wrapfig,times}
\usepackage{amsthm}
\usepackage{setspace}
\usepackage{subfigure}

\newcommand{\be}{\begin{equation}}
\newcommand{\ee}{ \end{equation}}
\newcommand{\ben}{\begin{eqnarray}}
\newcommand{\een}{\end{eqnarray}}

\begin{document}

\title{Intermediate statistics: addressing the Landau diamagnetism problem}

\author{André A. Marinho$^{1,2}$, Francisco A. Brito$^{2,4}$, G.M. Viswanathan$^{1,3}$, C.G. Bezerra$^{1}$ }

\affiliation{$^{1}$ Departamento de Física, Universidade Federal do Rio
Grande do Norte, 59078-900 Natal, RN, Brazil
\\
$^{2}$ Departamento de Física, Universidade Federal de
Campina Grande, 58109-970 Campina Grande, Paraíba, Brazil
\\
$^3$National Institute of Science and Technology
                   of Complex Systems,
                   Universidade Federal do Rio Grande do Norte,
                   Natal--RN, 59078-900, Brazil
\\
$^{4}$ Departamento de Física, Universidade Federal da Paraíba,
Caixa Postal 5008, 58051-970 João Pessoa, Paraíba, Brazil}
\date{\today}

\begin{abstract}
Quantum groups and quantum algebras have received considerable attention in the last decades because they are very useful as mathematical tools of research. Existing proposals for quantum groups have always suggested the idea of deforming a classical object. Motivated by the possibility of anyons in three dimensions ($d=3$), with important consequences to a wide
range of fields of physics, in the present work we investigate how the magnetization and other thermodynamic quantities, associated to the Landau diamagnetism problem, depend on the deforming parameter of two models with intermediate statistics: (i) $q$-fermions and (ii) $F$-anyons, and make {\it comparisons between both cases}. In particular, we extend the results from the literature for $q$-fermions by considering {\it second order terms} in the expansion of the grand partition function. Also, we find that for $F$-anyons statistics the magnetization shows a stronger response with respect to magnetic fields compared to magnetization for $q$-fermions statistics. This theoretical outcome may be experimentally verified for instance in superconductors, that are perfect diamagnetic materials with strong magnetic susceptibility, by adjusting impurities or pressure. The latter can be associated to the deforming parameter $q$.
\end{abstract}


\maketitle

\section{Introduction}

In recent years, several fields of physics such as cosmology and condensed matter have intensified researches
in the study of quantum groups and quantum algebras, mainly due to the large number of possible applications
\cite{lei,wil,che,ien,ler,fuc,bie,mac,chu,chu1,chu2,hou,pra,bou}, as well as rational field theories, fractional quantum Hall
effect, high-temperature (high-T$_c$) superconductors, noncommutative geometry, quantum theory of super algebras and so on
\cite{rpv,chai,col,sou,lav5}. All existing proposals for quantum groups suggest the idea of deforming a classical object, which
may be, for example, an algebraic group or a Lie group, in such way that is always possible to preserve the understanding of the various representations that objects can admit \cite{arik,jim}.

It is known from the literature a number of intermediate statistics
(or fractional statistics) works describing anyons (or quasifermions)
\cite{aro,sen,ach,ach1,per,lav,rov1,aba,aba1,aba3,aba4,nar2,shen,gli,lens,Morier,greiter,kwan},
which is one of versions of non-standard quantum statistics
\cite{gen,gre,pol}.  Let us recall the usual argument that tells us we
should restrict to bosons and fermions. We take two identical
particles described by the wavefunction $\psi(\vec{r_1},\vec{r_2})$.
Since the particles are identical, all probabilities must be the same
if the particles are exchanged \cite{wil}. This tells us that
$|\psi(\vec{r_1},\vec{r_2})|^2=|\psi(\vec{r_2},\vec{r_1})|^2$ so that,
upon exchange, the wavefunctions differ only by a global phase
\be\label{eq1}\psi(\vec{r_1},\vec{r_2})=\exp(i\pi\theta)\psi(\vec{r_2},\vec{r_1}).\ee
Eq.\ (1) provides the two familiar possibilities: bosons $(\theta=0)$
and fermions $(\theta=1)$. From a mathematical perspective, we can
normally state that for spatial dimensions $d\geq 3$, the exchange of
particles must be described by a representation of the {\it
  permutation group}. On the other hand, for dimension $d=2$,
exchanges are described by a representation of the \textit{braid
  group}. This latter case is the realm of the intermediate statistics
that usually describes anyons. However, recall that, in general,
fermions can obey a more general exclusion principle, such as the
Pauli-Haldane exclusion principle, which allows the existence of
anyons or fractional statistics in arbitrary $d$ dimensions
\cite{haldane}.

Indeed, studying fractional statistics in 3D or higher dimensions may well offer a way to generalize quantum statistics and explore novel quantum phases: (i) it may end up revealing new types of quasiparticles, analogous to anyons in 2D, with properties relevant to systems like topological insulators \cite{haldane}; (ii) it may help extend topological quantum computing by exploring how non-Abelian anyons or similar quasiparticles behave in higher dimensions, potentially leading to more robust quantum systems; and (iii) fractional statistics in higher dimensions might also provide generalizations that apply to systems beyond the usual fermionic and bosonic statistics. Finally, there is the hope that this approach might contribute towards strengthening the bridge between condensed matter physics on the one hand with high-dimensional theories in quantum field theory or string theory on the other hand, thereby offering new insights into particle interactions in those spaces.  We thus take it for granted that there is no {\it a priori} reason not to study fractional statistics in 3D. As a matter of the fact, Teo and Kane have shown the
possibility of anyons in $d=3$, with important consequences to a wide range of fields of physics, such as many-body condensed matter systems, quantum computation, and the fundamentals of quantum
mechanics \cite{teo}.

The aim of this work is to address the issues of fractional statistics
of (i) $q$-fermions and (ii) $F$-anyons ($F$-type systems). Although
the latter case does not obey an exclusion principle in general, the
former case obeys at least the Pauli exclusion principle. In
particular, we extend the results from the literature for $q$-fermions
by considering {\it second order terms} in the expansion of the grand
partition function. Furthermore, we will explore the analytical
results, and make {\it comparisons between both cases}. We thus focus
our attention on the study of the Landau diamagnetism problem
\cite{lan}, which continues to raise issues of great relevance
nowadays \cite{oze,omer,am1,am2}, mainly due to the inherent quantum nature of
the problem. It can also be used as a phenomenon to illustrate the
essential role of quantum mechanics on the surface and perimeter, the
dissipation of the statistical mechanics of non-equilibrium, and other
relevant physical problems. In the present study we show how the
magnetization, and other thermodynamic quantities, in the Landau
diamagnetism problem, depend on the deforming parameter of the two
scenarios considered ($q$-fermions and $F$-anyons). This paper is
organized as follows. In Sec.\ II we introduce the algebra and the
Jackson derivative for $q$-fermions. The algebra for $F$-anyons is
described in Sec.\ III. We develop the Landau diamagnetism problem
working with both types of $q$-deformation (and make comparisons
between them) in Sec.\ IV. Finally, in Sec.\ V we summarize the main
results obtained in this work and make our final comments.

\section{$q$-Fermions Algebra}
\label{alg}

In this section we shall focus our attention to the thermodynamics and statistical mechanics of fermions based on the $q$-deformed algebra. Let us start from the deformed anticommutation relations \cite{nar2},
\begin{equation} \label{eq2} ff^{\dagger} + qf^{\dagger}f = q^{-{N}},\end{equation}
\begin{equation}[{N},f] = -f, \qquad [{N},f^{\dagger}] = f^{\dagger} \qquad f^2 \neq 0.\end {equation}
Here $f$ and $f^{\dagger}$ are the deformed fermionic annihilation and creation operators and ${N}$ is the fermion
number operator, and $q$ is the {\it deformation parameter} of the model. We remark that this algebra cannot be reduced to the following
standard fermion oscillator algebra due to the relation $f^2\neq 0$,
\be ff^{\dagger} + f^{\dagger}f = 1.\ee

\noindent Therefore, in a theory based on this $q$-fermion algebra, the number operator $N$ eigenvalues are not restricted to the values
$n = 0, 1$ and the Pauli principle of exclusion is not satisfied. Bearing this in mind, we will introduce the following definition of fermion \textit{basic number} \cite{aba2,aba5,aba6,nar,nar1,nar3,vis,chai,lav1,erns,flo}
\begin{equation} \label{eq3} f_{i}^{\dagger}f_{i} = [N] = \frac{q^{-N}-(-1)^{N}q^{N}}{q+q^{-1}}.\end{equation}

\noindent The $q$-Fock space spanned by the orthonormalized eigenstates $|n \rangle$ is constructed according to
\begin{eqnarray} {|n\rangle} =\frac{(f^{\dagger})^{n}} {\sqrt{[n]!}}{|0\rangle},\qquad\qquad f{|0\rangle}=0.\end {eqnarray}
The actions of $f$, $f^{\dagger}$ and $N$ on the states $|n\rangle$ in the $q$-Fock space are known from the literature
\begin{equation} f^{\dagger}{|n\rangle} = (n+1)^{1/2} {|n+1\rangle},\qquad
f{|n\rangle} = (n)^{1/2} {|n-1\rangle},\qquad N{|n\rangle} = n{|n\rangle}.\end{equation}

For the purpose of calculating the $q$-deformed {\it fermionic} occupation number (from now on labeled $n_i^{(q)}$), we begin with the Hamiltonian of the deformed fermion oscillator,
\be\label{eq20} {{\cal H}}_{q} = \sum_{i}{(\epsilon_i-\mu)}{{N}},\ee
where $\mu$ is the chemical potential, $\epsilon_i$ is the energy of a particle in the state $i$ and
${N}$ is the fermion number operator. It should be noted that this Hamiltonian is deformed
and implicitly depends on the deformation parameter $q$, since the number operator is deformed by means of Eq.\ (\ref{eq3}).

\noindent The mean value of the $q$-deformed occupation number $n_{i}^{(q)}$ can be calculated and is given by
\ben [n_i]\label{eq4.1}\equiv \langle[n_i]\rangle = \frac{tr\Big[\exp(-\beta {{\cal H}}_{q} )
f_{i}^\dagger f_{i}\Big]}{\Xi},\een
where we apply the cyclic property of the trace \cite{ams,amg2} to obtain,
\ben [n_i] = \frac{\exp\Big[-\beta(\epsilon_i-\mu)\Big]tr\left\{\exp\Big[-\beta\displaystyle\sum_{n_j}
{(\epsilon_j-\mu)n_j f_{i}f_{i}^\dagger}\Big]\right\}}{tr\left\{\exp\Big[-\beta\displaystyle\sum_{n_j}{(\epsilon_j-\mu)n_j }
\Big]\right\}},\een
and taking into account the definition $f_i f_{i}^\dagger=[1+n_i]=q^{-n_{i}}-q[n_i]$, we get
\ben \label{eq5}[n_i] = \exp[-\beta(\epsilon_i-\mu)][1+n_i]\,\, \rightarrow \,\, \frac{[n_i]}
{[1+n_i]} = z\exp(-\beta\epsilon_i).\een
Thus,  $n_{i}^{(q)}$, with the positivity condition satisfied, is finally given by 
\be \label{eq7} n_{i}^{(q)}=\frac{1}{2|\ln{(q)}|}\left|\ln\left(\frac{|z^{-1}\exp(\beta\epsilon_i)-q^{-1}|}
{z^{-1}\exp(\beta\epsilon_i)+q}\right)\right|.\ee
Eq.\ (12) is the $q$-deformed fermionic distribution function for all $q$ in the interval $0<q<1$.
\subsection{Jackson derivatives for fermions}

Let us now discuss a little about a mathematical tool that will be applied later on in this work: the Jackson derivative (JD) \cite{jak}. In the last decades several proposals for extending the concept of calculus, with great potential of application for the study of deformed systems have appeared. Among these proposals we can cite many generalizations of the usual Leibniz derivative, such as the Jackson derivative, fractional derivatives and the Hausdorff derivative, which can be applied in a large set of functions. In particular, our interest in JD stems from the fact that it is the $q$-analog of the ordinary derivative. Unlike more recent deformations of the geometric and hypergeometric series, the theory of $q$-series has centuries of solid mathematical underpinning. Given this long mathematical tradition, the $q$-analogs have a far more solid theoretical basis in comparison to other deformations of the standard algebra.

The JD is obtained from the basic number definition and mathematical development presented above. Depending on a particular definition
we can obtain a JD which at the limit $q\to 1$ is reduced to ordinary derivative. In short, by considering the fermion \textit{basic number} Eq. (\ref{eq3}) we may transform the $q$-Fock space into the configuration space (Bargmann holomorphic representation) \cite{flo} as
\begin{eqnarray} \label{e32}f^{\dagger} = x,\qquad\qquad f = \partial_{x}^{(q)},\end{eqnarray}
where $\partial_{x}^{(q)}$ is the Jackson derivative for fermions \cite{nar,nar1,nar3}
\ben \label{e33} \partial_{x}^{(q)}g{(x)}=\frac {g{(q^{-1}x)}-g{(-qx)}}{x{(q+q^{-1})}}, \een
where $g(x)$ is an arbitrary function. Notice that unlike other definitions found in the literature for de JD, e.g. \cite{bie,mac,chai,aba,lav1}, in this case the Eq.\ (\ref{e33}) does not return the ordinary derivative in the limit $q\to1$ \cite{nar}.

\section{$F$-Anyons Algebra}
In this section we shall focus our attention to the thermodynamics and statistical mechanics
of $F$-anyons based on the q-deformed algebra. Let us start from the algebra defined as \cite{nar2}

\begin{equation} \label{eq4} f_{i}f_{i}^{\dagger}+q^{-1}f_{i}^{\dagger}f_{i} = q^{-N},\qquad 0\leq q\leq1,\end{equation}
with the commutation relations
\begin{equation}[{N},f_{i}] = -f_{i}, \qquad [{N},f_{i}^{\dagger}] =f _{i}^{\dagger}. \end {equation}
Since this algebra is not related to {\it basic numbers} \cite{ext}, the rules of ordinary derivatives prevail. We will also see that $F$-anyons obey the Pauli
exclusion principle, and that although $q$-fermions and $F$-anyons  depend on the same deforming parameter $q$,
{\it their respective $q$-deformed distributions are different} and returns to the ordinary distribution in the limit
$q\to 1$.

By introducing the operator $f_{i}^{\dagger}f_{i}={N}$, and assuming that the action in the Fock state is described by \cite{nar,nar1,nar3}, we have
\be {N}{|n_{i}\rangle} = \alpha_{n_{i}}{|n_{i}\rangle},\qquad {N}f_{i}^{\dagger} + q^{-1}f_{i}^{\dagger}{N}=f_{i}^{\dagger}q^{-N},\ee
where the eigenvalue depends on $n_{i}$ and the relation follows from Eq.\ (\ref{eq4}). We may set
\be f_{i}{|n_{i}\rangle}=C_{n_{i}}{|n_{i}-1\rangle}, \qquad f_{i}^{\dagger}{|n_{i}\rangle}=C^{'}_{n_{i}}{|n_{i}+1\rangle}, \ee
where the constants $C_{n_{i}}$, $C^{'}_{n_{i}}$ can be determined. As a consequence we obtain the recurrence relation,
\be\label{eq21} \alpha_{n_{i}+1}=q^{-n_{i}}-q^{-1}\alpha_{n_{i}}, \ee
where $\alpha_{n_{i}}$, with ${n_{i}}=0$, is defined as the ground state. Therefore,
\be \alpha_{{n_{i}}}=0,1,0,q^{-2},0,q^{-4}, \cdots\qquad =\frac{1-(-1)^{n_{i}}}{2}q^{-{{n_{i}}}+1}.\ee
The action of $f_{i}$ and $f_{i}^{\dagger}$ on the Fock states yields
\be f_{i}^{\dagger}{|0\rangle}=\sqrt{\alpha_1}{|1\rangle}={|1\rangle}, \qquad f_{i}^{\dagger}f_{i}^{\dagger}{|0\rangle}=
\sqrt{\alpha_1}\sqrt{\alpha_2}{|2\rangle}=0,\ee
such that the possible states are ${|0\rangle}, {|1\rangle}$ only.

For the purpose of calculating the $q$-deformed {\it anyonic} occupation number (from now on labeled $n_{i,q}$), let us consider the same Hamiltonian given by Eq.\ (\ref{eq20})
and represent the $q$-deformed occupation number in the form
\ben {n}=\frac{{\rm Tr}\Big[\exp(-\beta {{\cal H}}_{q}){N}\Big]}{\Xi}=\frac{{\rm Tr}\Big[\exp(-\beta {{\cal H}}_{q})
f_{i}^\dagger f_{i}\Big]}{\Xi}.\een
Proceeding as before, and using Eq.\ (\ref{eq21}), we find
\begin {equation}\label{eq22} \frac{1-(-1)^{n}}{2} = \left[\frac{q^{-1}}
{z^{-1}\exp(\beta\epsilon_i)+q^{-1}}\right].\end{equation}
We may rewrite this to obtain the $n_{i,q}$ in the same way as \cite{nar, nar1, nar3}
\be \label{eq23}n_{i,q}=\frac{2}{\pi}\arcsin\left[\sqrt{\frac{q^{-1}}{{z^{-1}\exp(\beta\epsilon_i)+q^{-1}}}}\right].\ee
Eq.\ (\ref{eq23}) is not written in a form that facilitates its application in the determination of thermodynamic quantities.
However, there is a simplification for the case of $F$-anyons systems. We recall that Fock states are reduced to $n=0, 1$
only. We also note that $\sin^2(n\pi/2)=0, 1$ and hence can be replaced by $n$ without losing generality. Thus, we arrive at the
simplified form
\be \label{eq24}n_{i,q}=\frac{q^{-1}}{{z^{-1}\exp(\beta\epsilon_i)+q^{-1}}}.\ee

\begin{figure}[!htb]
\centerline{
\includegraphics[{angle=90,height=6.5cm,angle=270,width=6.5cm}]{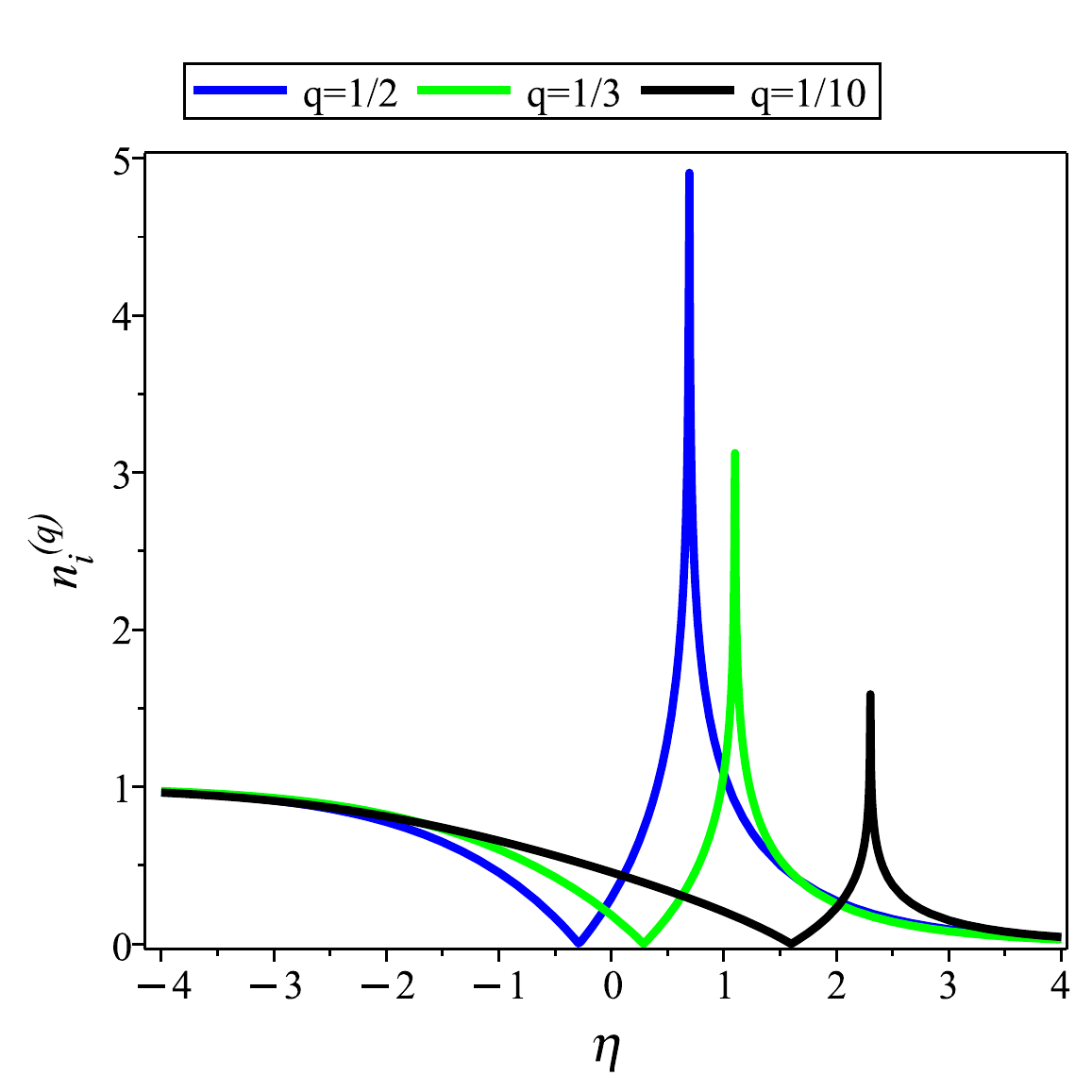}
\includegraphics[{angle=90,height=6.5cm,angle=270,width=6.5cm}]{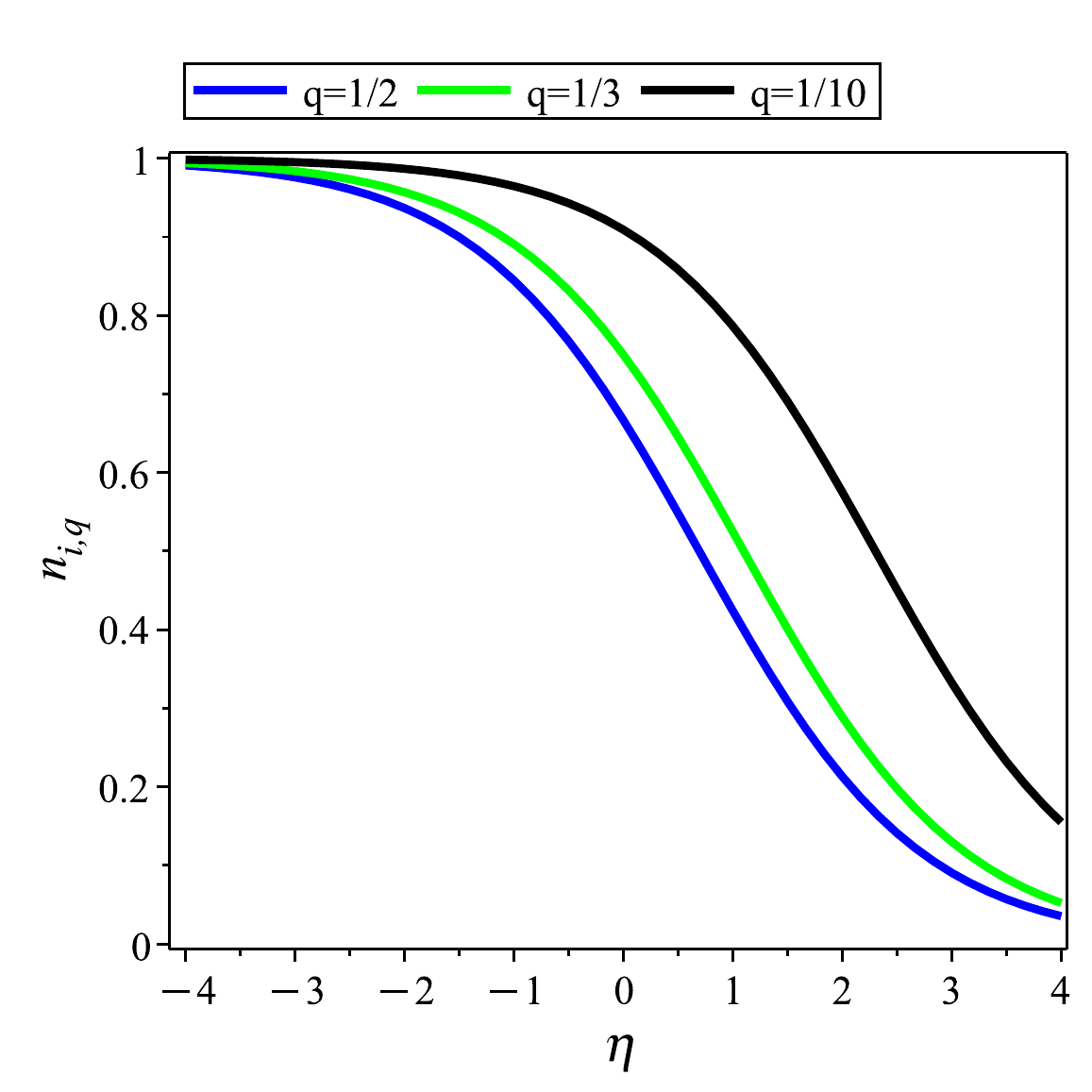}
\includegraphics[{angle=90,height=6.5cm,angle=270,width=6.5cm}]{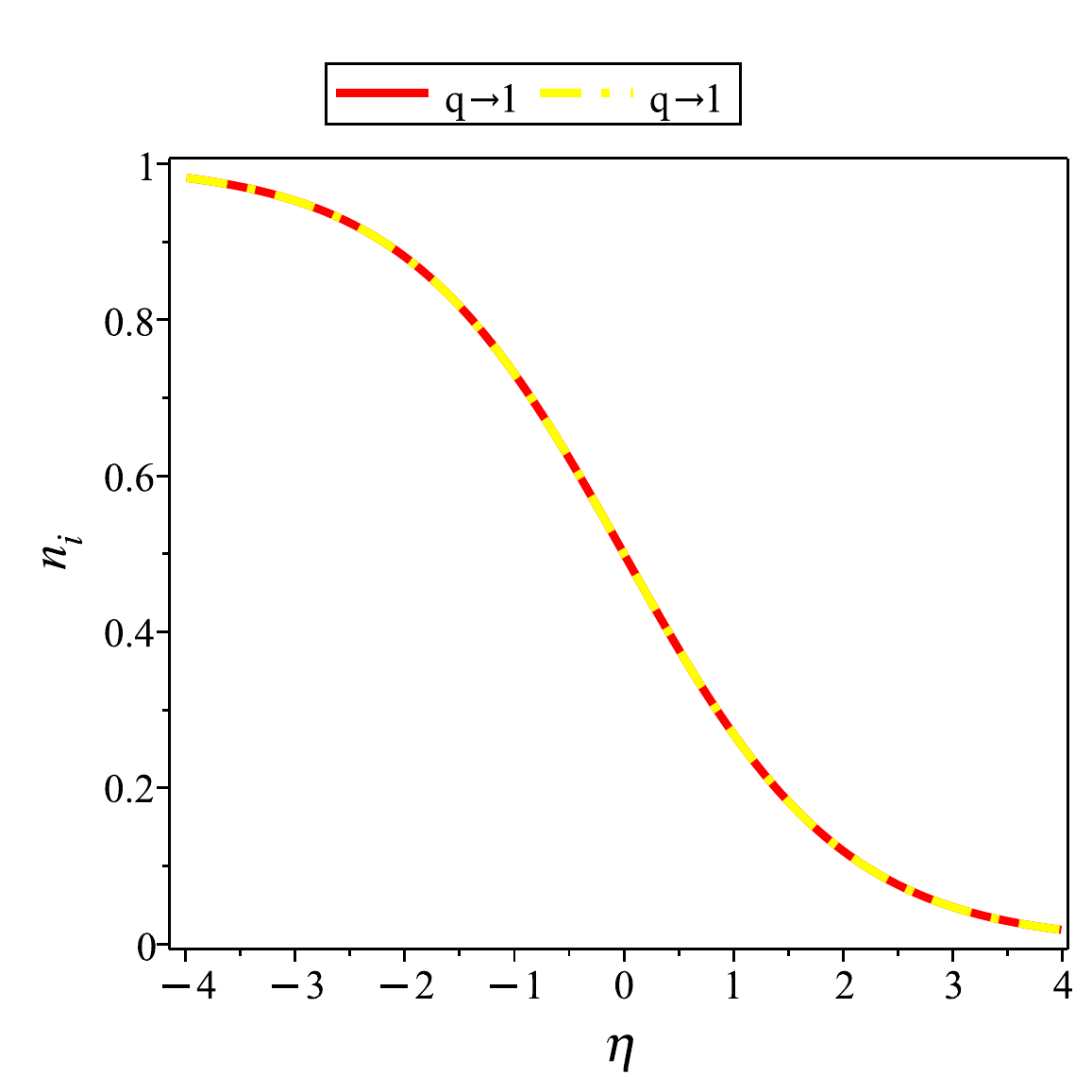}
}\caption{$q$-deformed fermionic distribution {$n_{i}^{(q)}$ (left) and $q$-deformed anyonic distribution $n_{i,q}$ (center) as a function of $\eta=\beta(\epsilon_i-\mu)$. Right: Overlap of $n_{i}^{(q)}$ and $n_{i,q}$ as $q\to 1$ .} Note how we recover the case of the fermion standard distribution function when $q\to 1$.}\label{graficos 1}
\end{figure}

We can now illustrate the behavior of $n_{i}^{(q)}$ ({\it fermionic case}) and $n_{i,q}$ ({\it anyonic case}) in Fig.\ (\ref{graficos 1}). It is quite clear from Fig.\ (\ref{graficos 1}) that $n_{i}^{(q)}$ and $n_{i,q}$ depend on the values of $q$, and when $q\to 1$ we recover the case of the fermion standard distribution function for finite values of temperature.

In the following we will apply the two cases of deformation given by Eqs.\ (\ref{eq7}) and (\ref{eq24})
to the Landau diamagnetism problem, obtaining generalized thermodynamic quantities.
\section{$q$-Deformed Landau Diamagnetism}
\label{qdld}

In order to address and explain the phenomenon of diamagnetism, we have to take into account the interaction between the applied external magnetic field
and the orbital motion of electrons. Disregarding the spin, the Hamiltonian of a particle of mass \textit{m} and charge \textit{e}
in the presence of a magnetic field \textbf{H} is given by \cite{sal}
\begin{equation} \label{eq8}{\cal H} = \frac {1}{2m} \left({\bf p}-\frac{e}{c}{\bf A}\right) ^2, \end{equation}
where \textbf{A} is the vector potential associated with the magnetic field \textbf{H} and \textit{c} is the speed of light
in CGS units.
\subsection{$q$-Fermions Application to the Landau diamagnetism problem}
\label{qfa}

We will describe the thermodynamic quantities for $q$-fermions algebra. Three approaches will be shown. Approach I: the deformation parameter $q$ is introduced by using JD over a non-deformed grand partition function, obtaining  $N_{1}^{(q)}$. Approach II: the deformation parameter $q$ is introduced via a $q$-deformed grand partition function and we apply the ordinary derivative, obtaining $N_{2}^{(q)}$. Finally, approach III: the deformation parameter $q$ is introduced trough both, applying JD in a $q$-deformed grand partition function, obtaining $N_{3}^{(q)}$. It is worth to remark again that we extend the results from the literature for $q$-fermions by considering {\it second order terms} in the expansion of the grand partition function.

\subsubsection{Approach I: application of the Jackson Derivative (JD) over a non-deformed grand partition function $\Xi$}

Let us start formalizing the statistical mechanical problem by using the logarithm of the grand partition function
$\Xi$ for the Landau diamagnetism problem, of the form \cite{sal}

\ben \label{e54} \ln\Xi = \frac{2eHL^3}{(2\pi) hc}\displaystyle\sum_{n}^{\infty}
\displaystyle\int_{-\infty}^{\infty}dk_{z}\left\{|\ln{\left[|1-z\exp(-\beta\epsilon_n)\right|]}|\right\}. \een
Here $k_{z} \in (-\infty,+\infty)$, $\epsilon_n=\frac{\hbar^{2} k_{z}^2}{2m}+\hbar\omega\left(n+\frac{1}{2}\right)$ and
$\omega=\frac{eH}{mc}$.

As we know the Landau diamagnetism problem is solved in the limit of high temperatures $(z\ll 1)$. Thus, we will carry out a Taylor
expansion until second order of Eq.\ (\ref{e54}), i.e.,

\ben \ln\Xi=\frac{zHC}{\sinh(\gamma)}+\frac{z^2HC\sqrt{2}}{4\sinh(2\gamma)},
\qquad \mbox{with}\qquad C=\frac{eL^3}{hc\lambda_T}. \een
Here, $\lambda_T={h}/\sqrt{2\pi m\kappa_B T}$ is the thermal wavelength, $\gamma=\beta\mu_{B}H$ and
$\mu_{B}=\frac{e\hbar}{2mc}$ is the Bohr magneton. The average occupation number from the algebra of the non-deformed quantum oscillator is given by
\begin{equation} \label{eq71}n_{i} = \frac{1}{z^{-1}\exp(\beta\epsilon_i)-1}.\end{equation}
We have that
\begin{eqnarray} N=\displaystyle\sum_{i}n_{i},
\end{eqnarray}
what is also true for the $q$-deformed case, i.e.,
\begin{eqnarray} N_{1}^{(q)} = \displaystyle\sum_{i}n_{i}^{(q)}.
\end{eqnarray}
\noindent We can determine the total number of particles $N$ from the logarithm of the grand partition function $\Xi$,
\begin{eqnarray}\label{eq91} N = z\frac{\partial}{\partial z}\ln{\Xi} = \displaystyle\sum_{i}n_{i}.\end{eqnarray}
\noindent However, the total number of particles in the formalism of the ${q}$-deformed oscillators $N_{1}^{(q)}$
can not be obtained by using usual thermodynamics. Therefore,
we have to establish a relationship between $N_{1}^{(q)}$ and $N$ performing an expansion
in $z\ll 1$ in Eqs.\ (\ref{eq7}) and (\ref{eq71}), i.e.
\begin{eqnarray} n_{i}^{(q)}=\frac{q+q^{-1}}{2|\ln(q)|}\;z \exp(-\beta\epsilon_{i})
\qquad \mbox{and} \qquad n_{i}=z\exp(-\beta\epsilon_{i}). \end{eqnarray}
\noindent So,
\begin{equation}\label{e43}n_{i}^{(q)}=\frac{q+q^{-1}}{2|\ln(q)|}\;n_{i}\end{equation}
\noindent and
\begin{eqnarray} \displaystyle\sum_{i}n_{i}^{(q)}=\frac{q+q^{-1}}{2|\ln(q)|}\;\displaystyle\sum_{i}n_{i}\;\;\; \Rightarrow \;\;\;
N_{1}^{(q)}=\frac{q+q^{-1}}{2|\ln(q)|}\; N.\end{eqnarray}
Thus, we determined the $q$-deformed total number of particles $N_{1}^{(q)}$
\begin{equation} N_{1}^{(q)} = z\;D_{z}^{(q)}\;\ln{\Xi} = \frac{q+q^{-1}}{2|\ln(q)|}\left[\frac{zHC}{\sinh(\gamma)}
+\frac{z^2HC\sqrt{2}}{2\sinh(2\gamma)}\right].\end{equation}
Here $D_{z}^{(q)}$ is the so-called deformed differential operator defined as \cite{jak}
\ben \label {e51} \frac{\partial}{\partial z}\rightarrow D_{z}^{(q)}, \qquad
D_{z}^{(q)}=\frac{q+q^{-1}}{2|\ln(q)|}\;\partial_{z}^{(q)}.\een

\noindent Let us now insert the JD in the internal energy. In order to do so, we need to make a change from its definition \cite{aba2,jak}
\ben \label{e64} U_{1}^{(q)} =-\frac{\partial{y_n}}{\partial\beta}\;D_{y_{n}}^{(q)} \ln\Xi,\qquad
\mbox{where $y_n=\exp(-\beta\epsilon_n)$}.\een
We obtain,
\ben U_{1}^{(q)}=\frac{(q+q^{-1})}{2|\ln(q)|}
\left[\frac{z\mu_{B}CH^2\coth(\gamma)}{\sinh(\gamma)}+\frac{z^2\mu_{B}CH^2\sqrt{2}\coth(2\gamma)}{2\sinh(2\gamma)}
\right].\een
Entropy and magnetization are determined from the thermodynamical derivative of the grand potential $\phi_{1}$.
However, as before, to apply JD we have to rewrite the derivative as follows \cite{aba2,jak}
\begin{eqnarray} S_{1}^{(q)} = - \frac{\partial y_n}{\partial T}D_{y_{n}}^{(q)} \phi_{1},
\qquad M_{1}^{(q)} = - \frac{\partial y_n}{\partial H}D_{y_{n}}^{(q)} \phi_{1},\qquad\mbox{and}\qquad \phi_{1}=-\frac{1}{\beta}\ln\Xi. \end{eqnarray}
The entropy is given by
\ben S_{1}^{(q)}=\kappa_B\frac{q+q^{-1}}{2|\ln(q)|}
\left[\frac{zHC(1+\gamma\coth(\gamma))}{\sinh(\gamma)}+\frac{z^2HC\sqrt{2}(1+2\gamma\coth(2\gamma))}{4\sinh(2\gamma)}
\right].\een
Now, applying the definition of the Langevin function and using $N_{1}^{(q)}$ to eliminate the chemical
potential, we get
\be \label{eq15}{\cal L}(\gamma) = \coth(\gamma)-\frac{1}{\gamma},\qquad {\cal L}(2\gamma) = \coth(2\gamma)-\frac{1}{2\gamma}.\ee
Finally, the magnetization is given by
\ben M_{1}^{(q)}=-\mu_BN_{1}^{(q)}
\left[\frac{2\sinh(2\gamma){\cal L}(\gamma)+z\sqrt{2}\sinh(\gamma){\cal L}(2\gamma)}
{2\sinh(2\gamma)+z\sqrt{2}\sinh(\gamma)}\right].\een
\subsubsection{Approach II: application of the Standard Derivative over a $q$-deformed grand partition function $\Xi^{(q)}$}

We will apply now the usual derivative to determine the $q$-deformed thermodynamic quantities for the Landau magnetism. As we have already mentioned,
the statistics for $q$-fermions and $q$-bosons are different. However, Lavagno \cite{lav1} showed that it is possible
to describe both statistics at the same time by inserting the term $\kappa=\pm 1$ (for $q$-bosons and $q$-fermions, respectively),
where in these cases at the limit of $q\to 1$ the JD return to the standard form found in the literature. On the other hand, we show that the definition of JD given by the Eq.\ (\ref{e33}) does not obey that limit, but in both cases when $q\neq 1$ there are the consequences of the acting of $q$-algebra. For this to be true, it is necessary to follow the definition made by Swamy \cite{nar2}. Thus, we define the logarithm of the deformed partition function as follows
\ben \label{e34}\ln\Xi^{(q)} = \frac{2eHL^3}{(2\pi) hc}\displaystyle\sum_{n}^{\infty}
\displaystyle\int_{-\infty}^{\infty}dk_{z}\left\{\frac{q+q^{-1}}{2|\ln(q)|}|\ln{\left[|1-z\exp(-\beta\epsilon_n)\right|]}|\right\}, \een

\noindent that after performing some calculations takes the form
\ben \label{eq10}\ln\Xi^{(q)}=\frac{q+q^{-1}}{2|\ln(q)|}\left[\frac{zHC}{\sinh(\gamma)}
+\frac{z^2HC\sqrt{2}}{4\sinh(2\gamma)}\right].\een

\noindent In the following, we obtain $q$-deformed thermodynamic quantities by applying ordinary derivatives. For example, the number of particles $N$ and the internal energy $U$,
\ben\label{eq11} N_{2}^{(q)}=z\frac{\partial}{\partial z}\ln{\Xi^{(q)}}=\frac{q+q^{-1}}{2|\ln(q)|}\left[\frac{zHC}{\sinh(\gamma)}
+\frac{z^2HC\sqrt{2}}{2\sinh(2\gamma)}\right],\een
\ben \label{eq12} U_{2}^{(q)}=-\frac{\partial}{\partial\beta}\ln\Xi^{(q)}=\frac{q+q^{-1}}{2|\ln(q)|}
\left[\frac{zC\mu_{B}H^2\coth(\gamma)}{\sinh(\gamma)}
+\frac{z^2C\mu_{B}H^2\sqrt{2}\coth(2\gamma)}{2\sinh(2\gamma)}\right].\een
Also, the grand potential $\phi_{2}^{(q)}$ is obtained as
\be \label{eq13}\phi_{2}^{(q)}=-\frac{1}{\beta}\ln\Xi^{(q)}=-\frac{q+q^{-1}}{2|\ln(q)|}\left[\frac{zHC}{\beta\sinh(\gamma)}
+\frac{z^2HC\sqrt{2}}{4\beta\sinh(2\gamma)}\right],\ee
from which we can determine the entropy
\ben \label{eq14}S_{2}^{(q)}=-\frac{\partial\phi^{(q)}}{\partial T}=\kappa_B\frac{q+q^{-1}}{2|\ln(q)|}
\left[\frac{zHC(1+\gamma\coth(\gamma))}{\sinh(\gamma)}+\frac{z^2HC\sqrt{2}(1+2\gamma\coth(2\gamma))}{4\sinh(2\gamma)}
\right].\een

\noindent Finally, deriving Eq.\ (\ref{eq13}) with respect to the magnetic field $H$, using $N_{2}^{(q)}$ to eliminate the chemical potential,
and inserting the Langevin functions given in Eq.\ (\ref{eq15}), we obtain the magnetization
\ben \label{eq16}M_{2}^{(q)}=-\frac{\partial\phi^{(q)}}{\partial H}=-\mu_BN_{2}^{(q)}
\left[\frac{2\sinh(2\gamma){\cal L}(\gamma)+z\sqrt{2}\sinh(\gamma){\cal L}(2\gamma)}
{2\sinh(2\gamma)+z\sqrt{2}\sinh(\gamma)}\right].\een
\subsubsection{Approach III: application of the Jackson Derivative (JD) over a $q$-deformed grand partition function $\Xi^{(q)}$}

In the previous cases we insert the deformation parameter $q$ through two distinct ways: one starting from the
insertion of the JD through a deformed derivative over a non-deformed grand partition function $\Xi$, and the other way around,
i.e., considering a $q$-deformed grand partition function $\Xi^{(q)}$ and using ordinary derivatives. Now, we will go through a third way: considering the $q$-deformed grand partition function $\Xi^{(q)}$ from Eq.\ (\ref{e34}) and applying JD defined by Eq.\ (\ref{e33}).

Therefore, for the number of particles $N$ and the internal energy $U$ we have

\ben N_{3}^{(q)}=z~\partial_{z}^{(q)}\ln{\Xi^{(q)}}=\frac{q+q^{-1}}{2|\ln(q)|}\left[\frac{zHC}{\sinh(\gamma)}+
\frac{z^2HC\sqrt{2}}{2\sinh(2\gamma)}\right]\een
and
\ben U_{3}^{(q)}=-\frac{\partial{y_n}}{\partial\beta}\;\partial_{y_n}^{(q)}\ln\Xi^{(q)}=\frac{q+q^{-1}}{2|\ln(q)|}
\left[\frac{zC\mu_{B}H^2\coth(\gamma)}{\sinh(\gamma)}+\frac{z^2C\mu_{B}H^2\sqrt{2}\coth(2\gamma)}{2\sinh(2\gamma)}\right].\een

\noindent The entropy is calculated by making use of the great potential given by Eq.\ (\ref{eq13})
\ben S_{3}^{(q)}=-\frac{\partial y_n}{\partial T}\partial_{y_n}^{(q)}\phi_{3}^{(q)}=\kappa_B\frac{zHC(q+q^{-1})}{2|\ln(q)|}
\left\{\frac{\Big[1+\gamma\coth(\gamma)\Big]}{\sinh(\gamma)}+
\frac{z\sqrt{2}\Big[1+2\gamma\coth(2\gamma)\Big]}{4\sinh(2\gamma)}\right\}.\een

\noindent As before, deriving Eq.\ (\ref{eq13}) with respect to the magnetic field $H$, using $N_{3}^{(q)}$ to eliminate the chemical potential,
and inserting the Langevin functions given in Eq.\ (\ref{eq15}), we obtain the magnetization
\ben M_{3}^{(q)}= - \frac{\partial y_n}{\partial H}\;\partial_{y_n}^{(q)} \phi_{3}^{(q)}=-\mu_BN_{3}^{(q)}
\left[\frac{2\sinh(2\gamma){\cal L}(\gamma)+z\sqrt{2}\sinh(\gamma){\cal L}(2\gamma)}
{2\sinh(2\gamma)+z\sqrt{2}\sinh(\gamma)}\right].\een

\begin{figure}[!htb]
\centerline{
\includegraphics[{angle=90,height=12.0cm,angle=270,width=12.0cm}]{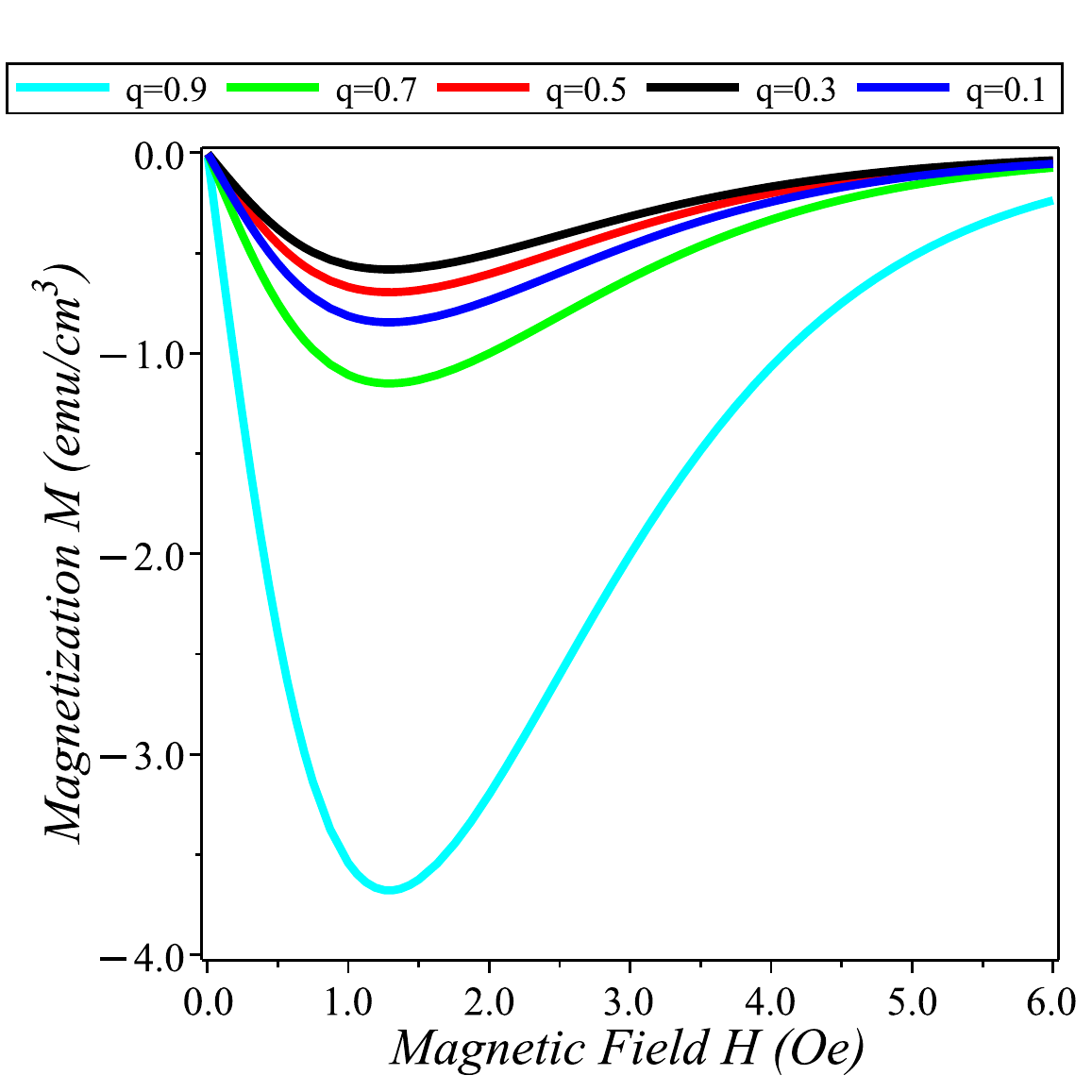}
}\caption{{$q$-deformed magnetization $M_{1}^{(q)}=M_{2}^{(q)}=M_{3}^{(q)}$ as a function of the magnetic field H
for different values of $q$ ($q$-deformed fermionic case).}}
\label{graficos 2}
\end{figure}

Therefore, it must be remarked that, even having followed three distinct ways of deforming the thermodynamic quantities, we show that for the case of
$q$-fermions in high temperature threshold the results are the same, i.e., $M_{1}^{(q)}=M_{2}^{(q)}=M_{3}^{(q)}$. Fig.\ \ref{graficos 2} shows magnetization M versus magnetic field H, for different values of the deforming parameter $q$. From Fig.\ \ref{graficos 2} is quite clear the dependence of the magnetization on the deforming parameter $q$, with $M_{1}^{(q)}=M_{2}^{(q)}=M_{3}^{(q)}$.

\subsection{$F$-Anyons application}
\label{faa}

Let us address now the Landau diamagnetism problem taking into account the intermediate statistics for $F$-anyons. The mathematical steps performed in the sequence will be the same for the $q$-fermions application. We start from Eq.\ (\ref{eq24}), taking the logarithm of the $q$-deformed grand partition function $\Xi_{q}$. Therefore,
\ben \label{eq27} \ln\Xi_{q} = \frac{2eHL^3}{hc (2\pi)}\displaystyle\sum_{n}^{\infty}
\displaystyle\int_{-\infty}^{\infty}k_{z}\ln[1+zq^{-1}\exp(-\beta\epsilon_n)]. \een
In the limit of high temperatures $(z\ll 1)$,

\ben \label{eq28}\ln\Xi_{q}=\frac{zHC}{q\sinh(\gamma)}-\frac{z^2HC\sqrt{2}}{4q^2\sinh(2\gamma)}
,\qquad \mbox{with}\qquad C=\frac{eL^3}{hc\lambda_T}.\een

\noindent From Eq.\ (\ref{eq28}) we obtain the number of particles $N_q$, the internal energy $U_q$ and
the grand potential $\phi_{q}$, i.e.,
\ben\label{eq29} N_{q}=z\frac{\partial}{\partial z}\ln{\Xi_{q}}=\frac{zHC}{q\sinh(\gamma)}
-\frac{z^2HC\sqrt{2}}{2q^2\sinh(2\gamma)},\een

\ben \label{eq30} U_{q}=-\frac{\partial}{\partial\beta}\ln\Xi_{q}=\frac{zC\mu_{B}H^2\coth(\gamma)}{q\sinh(\gamma)}
-\frac{z^2C\mu_{B}H^2\sqrt{2}\coth(2\gamma)}{2q^2\sinh(2\gamma)},\een
\noindent and
\be \label{eq31}\phi_{q}=-\frac{1}{\beta}\ln\Xi_{q}=-\left[\frac{zHC}{q\beta\sinh(\gamma)}
-\frac{z^2HC\sqrt{2}}{4q^2\beta\sinh(2\gamma)}\right].\ee

\noindent The entropy is determined from $\phi_{q}$,
\ben \label{eq32}S_{q}=-\frac{\partial\phi_{q}}{\partial T}=\kappa_B\left[\frac{zHC(1+\gamma\coth(\gamma))}{q\sinh(\gamma)}
-\frac{z^2HC\sqrt{2}(1+2\gamma\coth(2\gamma))}{4q^2\sinh(2\gamma)}\right].\een

\noindent Deriving Eq.\ (\ref{eq31}) with respect to the magnetic field $H$, using $N_{q}$ to eliminate the chemical potential, and inserting the Langevin functions given in Eq.\ (\ref{eq15}), we obtain the magnetization for the $F$-anyons case,
\ben \label{eq33}M_{q}=-\frac{\partial\phi_{q}}{\partial H}=-N_{q}\mu_B\left[\frac{2q\sinh(2\gamma){\cal L}(\gamma)+z\sqrt{2}\sinh(\gamma)
{\cal L}(2\gamma)}{2q\sinh(2\gamma)-z\sqrt{2}\sinh(\gamma)}\right].\een
\begin{figure}[!htb]
\centerline{
\includegraphics[{angle=90,height=12.0cm,angle=270,width=12.0cm}]{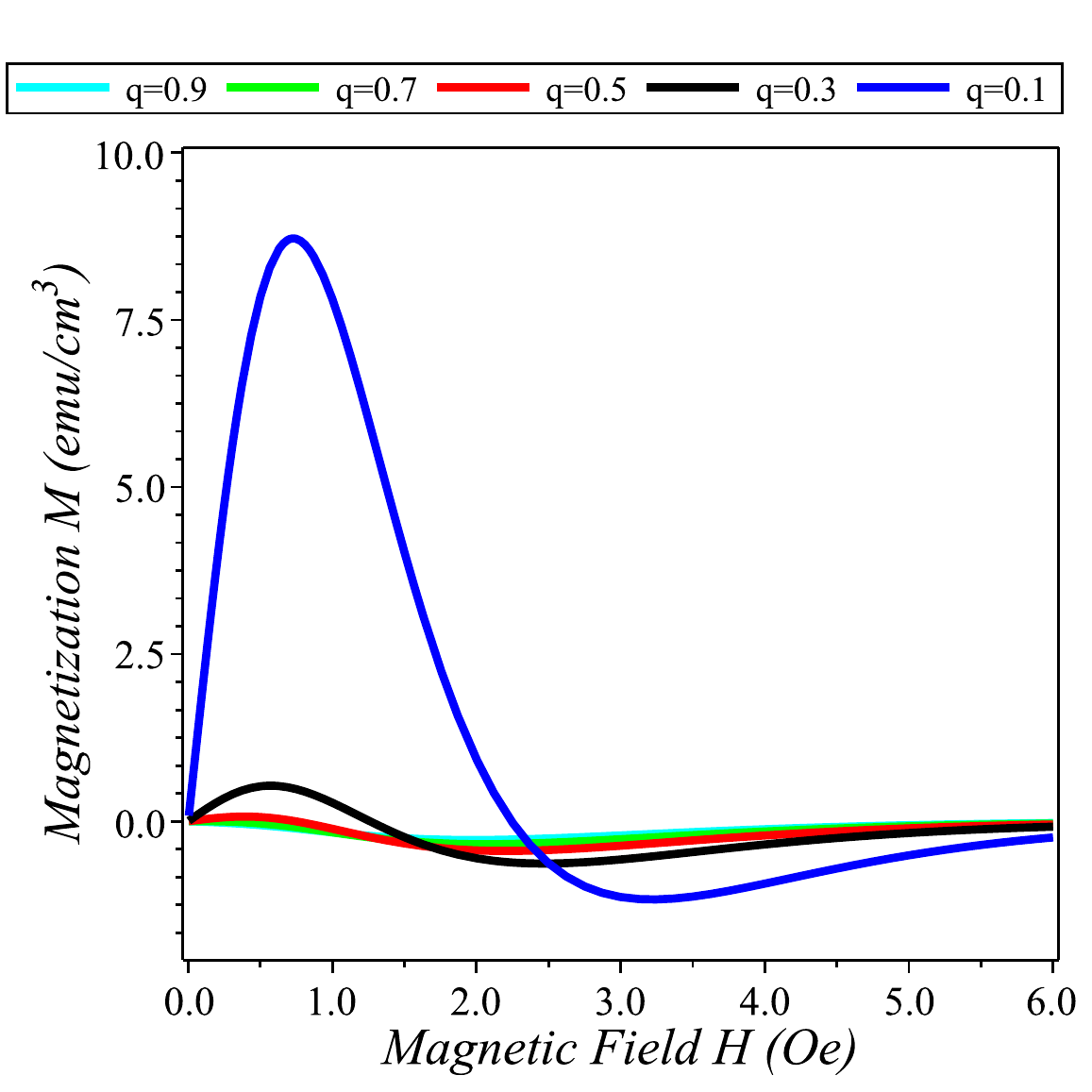}
}\caption{{$q$-deformed magnetization $M_{q}$ as a function of the magnetic field H
for different values of $q$ ($q$-deformed anyonic case).}}
\label{graficos 3}
\end{figure}
In Fig.\ (\ref{graficos 3}) it is shown the magnetization M as a function of the magnetic field H, for different values of the parameter $q$. It is clear comparing,
for the same values of $q$, Fig.\ (\ref{graficos 2}) ($q$-fermions) and Fig.\ (\ref{graficos 3}) ($F$-anyons), that both deformations present quite different behaviors. We observe that for $F$-anyons statistics, the magnetization shows a stronger response in relation to the magnetic field, compared to the magnetization for $q$-fermion statistics, in the same field range. These distinct behaviors can be used experimentally to distinguish both systems, for example in superconductors, which are perfect diamagnetic materials and have strong magnetic susceptibility, by inserting impurities or changes in pressure or temperature, as well as, the effects of disorder and stress in magnetic systems presenting the Nernst effect \cite{melo}. Other example is the modeling of thermal properties of a BiSbTe alloys \cite{marinho}. In that work a series of nanofilms were produced and then heat treated at different temperatures, thus modifying their properties. In fact, it has been showed in the literature that different values of the deformation parameter $q$ could be associated to different disorders \cite{marinho1,marinho3,marinho5,algin2024} and also to impurities or doping. Notice that the fact that our results for $q$-fermions and $F$-anyons are different, even for the same values of the deforming parameter $q$, opens new avenues in the modeling of different experimental systems. For example, we can calculate the thermodynamic quantities of the experimental samples before they are produced, saving material sample and lab time
\section{Conclusions}
\label{con}

We investigate generalized fermions through two $q$-deformed models: $q$-fermions and $F$-anyons, addressing the Landau magnetism problem. We recall that Pauli exclusion principle is valid for non-deformed models, as expected, it remains valid for $q$-deformed $F$-anyons. However, for the $q$-fermions model it has to be enforced by hand, explicitly or implicitly \cite{nar}. Regarding the $q$-fermions case, we remark that the results from the literature were extended by considering second order terms in the expansion of the grand partition function. In addition, three approaches were applied: (i) the deformation parameter $q$ is introduced by using JD over a non-deformed grand partition function; (ii) introducing the parameter $q$ via a $q$-deformed grand partition function and using ordinary derivatives; and finally (iii) introducing parameter $q$ through both, using JD and a $q$-deformed grand partition function. We show that all three approaches provide the same results for the $q$-fermions case. Regarding the $F$-anyons case, they only allow the application of the usual derivatives, because, as we have seen, their algebra does not have a defined basic number. Thus, the application of ordinary derivatives is straightforward to obtain thermodynamic quantities from a $q$-deformed grand partition function. Further related issues can be addressed in more general algebras such as hybrid algebras \cite{Marinho:2019zny}.

Furthermore, regardless the model, we have seen that the parameter $q$ modifies the statistics, and that in the limit $q\to 1$ we have
the standard fermionic statistics, as expected. Of course, the fact that the Landau diamagnetism problem is solved in the limit of
high temperatures at first order, causes the models to be very well tractable. However, different from what we find in the literature \cite{sal}, we have expanded
the problem up to second order to get better expressions that explicitly show the effect of the $q$-deformed statistics. It is important to remark that for $F$-anyons statistics, the magnetization, illustrated in Fig.\ 3, shows a more strong response with respect to the magnetic field than the magnetization for $q$-fermions statistics, in the same range of field. The results reported here may be verified experimentally, for instance, in superconductors by inserting impurities, or changes in pressure or temperature if one assumes these tuning quantities are related with the $q$-deformation parameter. Once disorder can affect different thermodynamic response functions \cite{cbezerra}, the approaches applied here may be considered for a number of physical systems. Finally, we hope that experimentalists are encouraged to investigate our model.

\section*{Data Availability}

The datasets generated during and/or analysed during the current study are available from the corresponding
author on reasonable request.

\section*{Acknowledgments}

AAM acknowledges support from PNPD/CAPES. GMV acknowledges support from CNPq (Grant no.\ 302414/2022-3). FAB acknowledges support from CNPq (Grant no.\ 309092/2022-1). CGB acknowledges support from CNPq (Grant no.\ 309495/2021-0).

\end{document}